\documentclass[aps, prb, twocolumn, reprint, superscriptaddress, longbibliography]{revtex4-2}

\usepackage{amsfonts, amsmath, bm, graphicx}

\def\vec#1{{\bm{#1}}}
\def\mat#1{{\hat{\vec{#1}}}}
\def\t#1{{\mathrm{#1}}}

\def\vp{{\varphi}}

\begin{document}
	
\title{Parametric excitation of nano-scale spin waves with controlled phase by an open-resonator-type transducer}

\author{Roman Verba} \email[corresponding author, e-mail: ]{verrv@ukr.net}
\affiliation{V. G. Baryakhtar Institute of Magnetism of the NAS of Ukraine, Kyiv 03142, Ukraine}

\author{Vladyslav Borynskyi}
\affiliation{V. G. Baryakhtar Institute of Magnetism of the NAS of Ukraine, Kyiv 03142, Ukraine}

\author{Vasil Tiberkevich}
\affiliation{Department of Physics, Oakland University, Rochester, Michigan 48309, USA}
	
\author{Andrei Slavin}
\affiliation{Department of Physics, Oakland University, Rochester, Michigan 48309, USA}
	
\date\today
	
\begin{abstract}
	Parametric excitation is one of the most effective methods for the excitation of nano-scale spin waves (SW) (or magnons) in magnetic films and nano-waveguides, as this method  removes  restrictions on the excited wave wavelength imposed by a finite width of the excitation transducer. Especially efficient parametric excitation of SW is achieved if the SW excitation is based on the  effect of voltage-controlled magnetic anisotropy (VCMA). However, in a traditional parametric excitation process, the excited propagating waves have an \emph{arbitrary} phase, which prevents the use of such waves in signal processing applications. Here, we demonstrate that the use of an VCMA transducer in the form of an \emph{open resonator} with controlled \emph{partial reflections} of waves at the boundaries allows one to excite SW having only two allowed phase states, controlled by the phase of the parametric pumping and reflections at the transducer boundaries. These reflections, also, lead to a reduced  value of the  parametric excitation threshold. In addition, micromagnetic simulations reveal the possibility to tune the phase distribution of the excited magnons using static voltage, which could be useful for applications in probabilistic computing.
\end{abstract}

\maketitle
	
	
\section{Introduction}

Magnonics -- the field considering the use of spin waves (SW) as information carriers -- offers intriguing perspectives for novel Boolean, and, especially, beyond-Boolean computing systems \cite{Pirro_NatRevMat2021, Chumak_IEEETM2022, Flebus_JPCM2024, Finocchio_NanoFut2024, Papp_NatComm2021, Wang_PRAppl2024}. A unique combination of features, inherent to SW, such as small wavelengths, high frequencies in gigahertz-sub-terahertz range, versatile tunability of the dispersion relations,  non-reciprocity, and efficient nonlinear and parametric processes, distinguish SW among other candidates for wave-based computing. In recent years, researchers achieved several milestones in the development and demonstration of key elements of magnonic circuits, including various logic gates \cite{Fischer_APL2017, Ustinov_IEEEMagLett2019, Wang_NE2020, Talmelli_SciAdv2020, Zenbaa_SciAdv2025}, magnonic neurons \cite{Fripp_APL2023, Fripp_APL2026, Breitbach_ArXiv2026}, auxiliary elements like amplifiers, repeaters, amplitude and phase normalizers \cite{Merbouche_NatComm2024, Nikolaev_SR2025, Wang_NC2024, Mahmoud_IEEETCSI2021, Verba_PRAppl2019a}, directional couplers \cite{Wang_SciAdv2018, Wang_NatComm2020}, etc., as well as successful demonstration of a magnonic integrated circuit \cite{Guo_ArXiv2026}.

Still, there are a lot of investigations and technological steps required for magnonic devices to become a market product, one of which is the transition to short exchange-dominated SW. Indeed, most of the demonstrations of magnonic circuit elements have been done for relatively long dipolar or dipole-exchange spin waves, which are relatively easy to access using traditional experimental techniques. However, reaching ultimate processing speed and energy efficiency requires utilization of short (submicron) spin  waves, which also leads to a reduction of the device footprint \cite{Wang_PRAppl2024}. Excitation of exchange-dominated spin waves is not a trivial task. Among different approaches --  using periodic magnetic structures \cite{Liu_NatComm2018}, tapered waveguides \cite{Demidov_APL2011} or other linear wavenumber-conversion mechanisms \cite{Che_NatComm2020}, spin-orbit torque emitters \cite{Demidov_NatComm2016}, SW emission via movement of a magnetization texture \cite{Sluka_NatNano2019} -- the parametric excitations \cite{Gurevich_Book1996, Bracher_PRep2017} looks the most attractive. Indeed, the parametric excitation does not impose any direct restrictions on the SW wavelength relative to the transducer width, allowing to  vary the wavelength in a wide range just by tuning  of the excitation frequency. An inherent problem of a weak parametric coupling efficiency for exchange-dominated SW due to their vanishing precession ellipticity is successfully solved if, instead of a microwave magnetic field pumping, one utilizes a pumping in the form  microwave modulation of the magnetic film anisotropy, which could be realized via the effect of voltage-controlled magnetic anisotropy \cite{Verba_PRAppl2017} or/and the magnetoelastic interaction \cite{Lisenkov_PRB2019}. 

A standard (first-order) parametric excitation process is a process of splitting of one quantum of the microwave pumping field into two quanta of SW (magnons) \cite{Gurevich_Book1996, Bracher_PRep2017}. In this process, only the sum of phases $(\vp_1+\vp_2)$ of the excited SWs is fixed by the pumping phase $\vp_p$, while the phase difference of the excited SWs $(\vp_1-\vp_2)$ remains arbitrary \cite{Gurevich_Book1996}. This is a nice feature for the SW amplification, as it allows to amplify SW independently of their phase,  and the parametric amplification of SW has been widely used in sub-millimeter magnetic samples in the past \cite{Kolodin_PRL1998, Melkov_JAP2001}, as well as it has been successfully realized in magnetic nano-waveguides recently \cite{Nikolaev_SR2025}.

However,  in the parametric excitation process of SW, this phase insensitivity results in a random phase of the excited magnons, determined by instant thermal magnon state at the initial moment of the pumping application. Thus, in the standard signal processing applications of the SW excited excited by traditional methods and  having a random phase can not be used. 

One known solution of this problem is the use of the so-called non-adiabatic parametric pumping \cite{Melkov_PRE2001}, when the spatial spectrum of the pumping has a substantial component at the wavenumber $2k$, where $k$ is the wavenumber of the excited SW. The interaction of $2k$ component of the pumping with the excited SW results in the phase dependence of the parametric interaction efficiency \cite{Melkov_PRE2001, Serga_JAP2003},  and, eventually, leads to the possibility of only two phase states of the excited SW. Realization of the non-adiabatic pumping requires either a very small size of the pumping localization $L_p$ ($k L_p <1$), or utilization of a spatially non-uniform (spatially modulated with $\pi/k$ period) pumping. The first approach leads to an enormous increase of the required pumping amplitude, which scales as $1/L_p$ with the pumping localization, and, also, creates additional problems of the impedance matching of a narrow microwave transducer with an external pumping circuit. The second approach makes the parametric excitation of SW  $k$-selective, thus killing the main advantage of the parametric excitation process. Somewhat better performance has been recently achieved by a combined approach -- inclusion of a local magnetic inhomogeneity with the size $l<k^{-1}$ into the pumping region \cite{Lentfert_APL2026}; understanding stability and tunability of such approach requires further investigation.

In this work, we propose an alternative route to impose an additional phase relation between the excited  parametric magnons, and, thus, to determine the possible phase states of the excited SW. This is achieved by placing an additional VCMA-controlled gate subjected to a DC voltage nearby the VCMA parametric transducer. DC voltage changes the local anisotropy under the gate, creating a partial reflection of the parametrically excited SW from it, so that the whole excitation structure resembles an open resonator. In addition to the determination of the phase of the excited SW,  the proposed route, also, leads to the reduction of the parametric excitation threshold in up to two times in the case of the full reflection of the magnon from DC gate.

The paper is organized as follows. Section \ref{s:idea} describes the main idea and the system under consideration. Analytical theory for the determination of the excitation threshold and the phase of the excited SW are presented in the Sec.~\ref{s:theory}. Analytical  prediction are verified using micromagnetic simulations, results of which are presented  in the Sec.~\ref{s:sim}. Finally, the conclusions are given in the Sec.~\ref{s:conclusion}.


\section{Principle of parametric excitation by an open resonator type transducer}\label{s:idea}

The standard first-order parametric process is the splitting of a microwave pumping photon at a frequency $\omega_p$ (or quantum of effective pumping field, if it is created via magnetoelectric or magnetoelastic effects) into two quanta of SW (magnons) at frequencies $\omega_1$ and $\omega_2$. This process obeys the energy conservation rule, $\omega_p = \omega_1 + \omega_2$, and the momentum conservation rule, $\vec k_p = \vec k_1 + \vec k_2$, which, in the case of adiabatic pumping (quasi-uniform pumping with the localization length $L_p \gg 1/|\vec k_{1,2}|$), is transformed into $\vec k_2 = -\vec k_1$. If spin wave spectrum is reciprocal, $\omega_\vec k = \omega_{-\vec k}$, the pumping excites two counter-propagating SW at the half pumping frequency, $\omega_{1,2} = \omega_p/2$ (for the case of nonreciprocal spectrum, see, e.g., \cite{Melkov_APL2006}). 

In addition, the parametric interaction implies the phase relation between the pumping and parametric magnons, $\vp_1+\vp_2 = \vp_p + \pi/2 + 2\pi n$, $n \in \mathbb{Z}$ \cite{Gurevich_Book1996} (in the equation which defines phase, there could be a different constant instead of $\pi/2$; this depends on the type of pumping). At the same time, the difference of phases of the parametric magnons is not fixed, making the values of these phases arbitrary. The exception is if the parametric magnons are quanta of the same SW, which is possible if $\vec k_1 = \vec k_2 = 0$, or in the case of standing SW modes of a magnetic nanodot. Then, one gets $\vp_1 = \vp_p/2 + \pi/4 + \pi n$, which is the relation, standard for parametric oscillator defining two possible distinct phase states of the excited oscillations.

In order to define the allowed phase states of the parametric magnons, one needs to introduce an additional phase relation. Since the pumping and magnons have different frequencies, there are no ways to relate the pumping phase with the phase of one of the magnons. In the case of  a non-adiabatic pumping, such an additional relation comes from the interaction with $2\vec k$ spectral component of the pumping $b_{2k}$, which implies a phase relation between $b_{2k}$ phase and $\vp_1$ (and a complementary relation between $b_{-2k}$ and $\vp_2$) \cite{Melkov_PRE2001}. 

In this work, we propose a different approach --  introducing a phase relation between the parametric magnons themselves. This is done using a partial reflection of one of the excited parametric magnons (for definiteness, magnon 2 with $-\vec k$ wave vector) from the DC gate. Upon the reflection, the magnon 2 changes its wave vector to $+\vec k$, thus becoming nothing else, but the magnon 1. Thus, we get an additional phase relation between $\vp_1$ and $\vp_2$. 

\begin{figure}
	\includegraphics[width=\columnwidth]{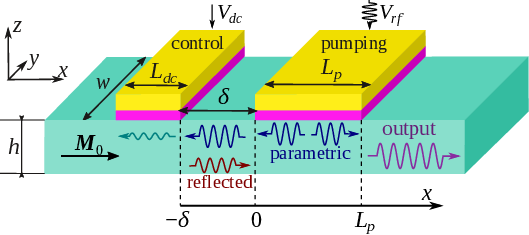}
	\caption{A sketch of the open resonator type transducer for the parametric excitation of SW. Ferromagnetic waveguide of the thickness $h$ and width $w$ is subjected to parametric pumping of the localization length $L_p$, created by microwave voltage $V_{rf}$ applied to metal-dielectric transducer. The pumping creates two counter-propagating parametric magnons; one of them is partially reflected from the control gate, subjected to a DC voltage. The interaction of the reflected magnon with the parametric magnon and the pumping results, finally, in the formation of a SW with a defined phase.}\label{f:sketch}
\end{figure}

A sketch of a system, which realizes the described concept, is shown in Fig.~\ref{f:sketch}. Here, we show a system utilizing VCMA parametric excitation, but the concept is valid for the parametric pumping of any other origin. The pumping metal-dielectric gate is subjected to microwave voltage $V_{rf}$, which results in the microwave modulation of perpendicular magnetic anisotropy under the gate via the  VCMA effect \cite{Miwa_JPDAP2018}. The pumping excites two counter-propagating magnons. The second magnon, propagating to the left (with a negative wavenumber) is reflected form the additional control gate, which is subjected to DC voltage leading to the shift of the SW dispersion under this gate. The reflection is described by the relation $a_1(-\delta) = R a_2(-\delta)$, where $a_j$ are complex SW amplitudes, and $R$ is the reflection coefficient, which, in general, could, also, be complex \cite{Verba_PRB2020bound}. Accounting for the phase accumulation acquired during the SW propagation from the pumping gate to the control gate and back (the separation between the gates is needed only for their electrical isolation) leads to the phase relation between the reflected wave (wave 1) and the wave 2 at the edge of the parametric gate is $\vp_1 = \vp_2 + \psi$, where $\psi = \mathrm{Arg}[R] + 2k\delta$. Taking into account this relation and the above mentioned phase relation imposed  by the parametric pumping, we find the possible phase states of the wave 1, which is the output wave (propagating to the right) in our transducer design:
	
\begin{equation}\label{e:phi-1-idea}
		\vp_1 = \frac{\vp_p + \psi}{2} + \frac{\pi}{4} + \pi n \,,
	\end{equation}
i.e., there are only two distinct phase states ($n = 0$ and $n=1$) -- similar to the case of a parametric oscillator. Of course, a small reflection coefficient and the influence of thermal fluctuations can smear the phase state of the output magnon.  This issue is addressed in the following section~\ref{s:sim} where  micromagnetic simulations are described.


\section{Theory}\label{s:theory}

In this section, we derive analytical description of the parametric excitation by an open-resonator-type transducer, which allows us to find the excitation threshold, as well as to confirm the above conclusion about the phase selectivity in the parametric excitation process. We start from standard equations,  describing evolution of amplitudes of the  SW envelopes $a_1$ and $a_2$ under the action of adiabatic parametric pumping \cite{Kalinikos_IEEE_TM1997}. For the stationary dynamics (i.e. $\partial_t = 0$) in a one-dimensional system (waveguide), these equations have the form:    
	\begin{equation}\label{e:Blomb}
		\begin{split}
			(&v\partial_x + \Gamma) a_1  = i V b_p (x) a_2^* \,, \\
			(&-v\partial_x + \Gamma) a_2^*   = -i V^* b_p^*(x) a_1 \ .
		\end{split}
	\end{equation}
Here, $v$ is the SW group velocity, $\Gamma$ is the damping rate, $V$ is the parametric coupling efficiency, which depends on the nature of the pumping (e.g. microwave magnetic field, modulated anisotropy, etc. ...) and a vector structure of a SW  (ellipticity of the magnetization precession), and $b_p(x)$ is a complex amplitude of the  effective pumping field at the double SW frequency, $\omega_p = 2\omega_k$. The pumping field is assumed to be piecewise constant, $b_p(x) = b_p$ under the pumping gate, and zero elsewhere, which is a natural assumption for the case of VCMA pumping. Equations (\ref{e:Blomb}) use the convention, that the parametric wave 1 is propagating in the positive ($+x$) direction, while the wave 2  is propagating in the negative  ($-x$) direction.

For the calculation of the excitation threshold, we use the same approach as Kalinikos et al. \cite{Kalinikos_IEEE_TM1997}. Namely, we consider formal incoming wave, which in our case is the wave 2 incident from $+\infty$ direction,  with arbitrary amplitude $A_2$ and phase $\vp_2$. This gives the first boundary condition at the right edge of the pumping gate (at $x = L_p$):
	\begin{equation}
		a_2(L_p) = A_2 e^{i \vp_2}  \ .
	\end{equation}
The second boundary condition relates the amplitudes of the magnons at the left boundary of the pumping gate. Accounting for the reflection from the control gate with a coefficient $R$, propagation losses and phase accumulation on the path from the pumping gate to the control one and back, we get :
	\begin{equation}
		a_1(0) = R a_2(0) e^{2ik\delta} e^{-2\Gamma \delta /v} = \tilde R e^{i\psi} a_2(0) \ .
	\end{equation}
Here, we define $\tilde R = |R| e^{-2\Gamma \delta/v}$ and $\psi = 2k\delta + \mathrm{Arg}[R]$.

First, let us consider the conservative approximation (setting the damping rate to zero $\Gamma \to 0$), as it allows for an easy phase analysis. 

General solution of Eq.~(\ref{e:Blomb}) then is
	\begin{equation}
		\begin{split}
			a_1(x)& = C_1 \cos[\kappa x] + i e^{i\vp_p} C_2 \sin[\kappa x] \,, \\
			a_2^*(x)& = C_2 \cos[\kappa x] + i e^{-i\vp_p} C_1 \sin[\kappa x]\,,
		\end{split}
	\end{equation}
where $\kappa = |V b_p|/v$. Applying the above mentioned boundary conditions, we get the amplitude of the output spin wave, which in our case is the wave 1 amplitude at the right edge of the pumping gate:
	\begin{equation}\label{e:a1-out-conserv}
		a_1(L_p) = A_2 \frac{\tilde R e^{i(\vp_2 + \psi)} + (i/2) (1+\tilde R^2) \sin[2\kappa L_p] e^{i(\vp_p - \vp_2)}}{\cos^2[\kappa L_p] - \tilde R^2 \sin^2[\kappa L_p]} \ .
	\end{equation}
The parametric excitation threshold is defined from the condition $|a_1(L_p)|/A_2 \to \infty$, which, physically, means that  any incoming thermal wave is infinitely amplified, i.e. becomes instable under the parametric pumping. This condition is $1-\tilde R \tan[\kappa L_p] = 0$, or, in the initial variables:
	\begin{equation}\label{e:th-conserv}
		|V b_\t{th}| = \frac{v}{L_p} \left(\frac{\pi}2 - \arctan \tilde R \right) \ .
	\end{equation}
In the limit $\tilde R \to 0$, this equation is reduced to a well-known condition for a standard adiabatic parametric excitation \cite{Kalinikos_IEEE_TM1997, Melkov_PRE2001}. It is clear, that the reflection from the control gate reduces the excitation threshold, with the maximum reduction of two times achieved in the case of full reflection at the boundary $\tilde R = 1$. This threshold reduction effect is explained by the reduction of the total radiative losses -- the wave 2 (propagating in $-x$ direction) is partially or fully suppressed outside the region under the pumping transducer. 
 
In the case of a full reflection, $\tilde R = 1$, Eq.~(\ref{e:a1-out-conserv}) allows for an easy phase analysis near the threshold. In this case, the numerator of Eq.~(\ref{e:a1-out-conserv}) is transformed to:
	\begin{equation*}
		2 \cos \left[\frac{\vp_p-\psi}{2} - \vp_2 + \frac{\pi}{4}\right] \exp\left[i\left(\frac{\psi + \vp_p}{2} + \frac{\pi}{4}\right)\right] \ .
	\end{equation*}
It is clear, that the output wave can have only two phases, depending on the cosine sign, and the values of these phases are exactly the same, as predicted above by the qualitative considerations, Eq.~(\ref{e:phi-1-idea}).

The full solution, accounting for the damping, is too cumbersome, and we present here only the condition of the parametric excitation, which is given by the following  implicit equation:
	\begin{equation}\label{e:th-full}
		\kappa v \cos[\kappa L_p] + \left(\Gamma - \tilde R |V b_\t{th}| \sin[\kappa L_p] \right) = 0 \,,
	\end{equation}
where 
	\begin{equation}
		\kappa = \frac{1}{v} \sqrt{|V b_\t{th}|^2 - \Gamma^2} \ .
	\end{equation}
The structure of Eq.~(\ref{e:th-full}) clearly demonstrates the effect of the reflection: reflection works against the damping, and reduces the threshold of parametric excitation. 

\begin{figure}
	\includegraphics[width=\columnwidth]{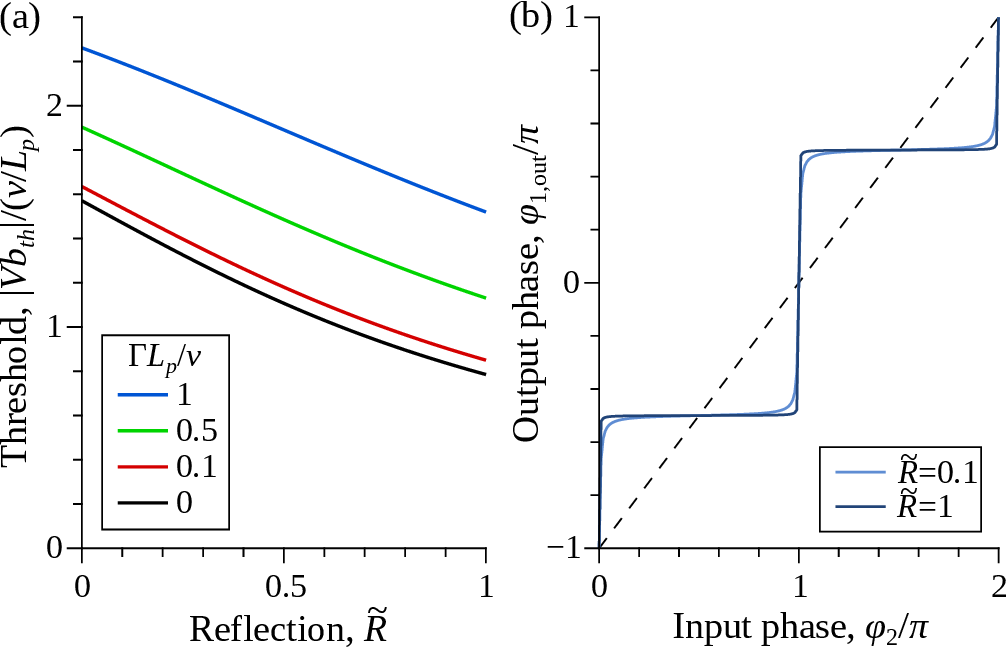}
	\caption{(a) The dependence of the threshold  of parametric excitation on the reflection coefficient for different SW damping rates (measured relative to the radiative losses); (b) Phase characteristic -- output SW phase vs input SW phase close to the excitation threshold (at the pumping level $b_p = 0.99 b_\t{th}$) for different reflection rates.  The ratio of the damping rate to the radiative losses is $\Gamma L_p / v = 1$. Dashed line shows the case  $\tilde R = 0$ (standard pumping by a single parametric gate).  The pumping phase is set to $\vp_p = \pi/2$, while the phase accumulation $\psi = 0$.}\label{f:th}
\end{figure}


The dependence of the parametric excitation threshold on the reflection coefficient, calculated using Eq.~(\ref{e:th-full}), is shown in Fig.~\ref{f:th}(a) for different damping rates $\gamma$ (normalized by the  radiative losses proportional to $v/L_p$). It is clear that the reflection reduces the excitation threshold in all the cases, and this reduction (if measured relative to the radiative losses as it is done in Fig.~\ref{f:th}(a)) is weakly dependent on the damping rate. This observation is easy to understand, as the reflection effect simply results in the reduction of the total radiative losses.

In Fig.~\ref{f:th}(b), we present the phase of the  output SW as a function of the input SW phase, which is obtained from the full solution of Eqs.~(\ref{e:Blomb}). Since at the threshold the solution is divergent, we set the pumping value at the level of $b_p = 0.99 b_\t{th}$. The resulting phase characteristics resemble the ``phase stabilization characteristics'', which is realized in the case of a non-adiabatic parametric pumping \cite{Verba_PRAppl2019a}. There are, in fact, only two possible phase states of the output SW,realization of one of which depend on the phase of the input SW. For the parametric excitation, where input wave is a thermal magnon having random-phase, this characteristics  means only two possible phases of the excited wave. It is worth noting, that phase selectivity arises even in the case of relatively small reflection coefficients, and is present independently of the damping magnitude (in particular, Fig.~\ref{f:th}(b) is plotted for relatively large damping, comparable to the radiative losses, $\Gamma L_p / v = 1$). In fact, when approaching the threshold even closer, all the phase characteristic tend to look as ideal step-like curves. However, in a  real system with thermal noise, smaller reflection coefficients leads to a larger phase dispersion, as it is demonstrated in the following section.


\section{Micromagnetic simulations}\label{s:sim}

In order to verify our theoretical conclusions and uncover other features of the parametric excitation of waves by an open-resonator-type transducer  we performed a series of micromagnetic simulations using MuMax3 solver \cite{Vansteenkiste_AIPAdv2014}. The simulated structure is shown in Fig.~\ref{f:sketch}. We consider CoFeB ferromagnetic waveguide of the thickness $h = 2\,$nm, width $w = 40\,$nm in zero magnetic bias field. Material parameters are: saturation magnetization $M_s = 800\,\t{kA}/\t{m}$, exchange stiffness $A_\t{ex} = 15\,\t{pJ}/\t{m}$, perpendicular surface anisotropy $K_s = 0.55\,\t{mJ}/\t{m}^2$ which is accounted for as the effective volume anisotropy with the constant $K_u = 275\,\t{kJ}/\t{m}^3$, effective Gilbert damping constant, accounting for both the Gilbert losses and nonuniform broadening contribution, is $\alpha_G = 0.02$. The VCMA coefficient, which relates surface anisotropy to the applied electric field, $K_s = \beta E$, is $\beta = 90\,\t{fJ}/(\t{V}\cdot\t{m})$, as reported for CoFeB/Mg/MgO structures in \cite{Li_APL2017}. Shape anisotropy and nanoscale sizes of the waveguide ensures its uniform static magnetization along the waveguide axis in a zero bias magnetic field (the waveguide thickness is chosen to be above the perpendicular-to-in-plane transition). 

Both the excitation and control gates were chosen to be of the length $L_p = L_\t{dc} = 500\,$nm, the distance between them was $\delta = 100\,$nm. The whole simulated length of the waveguide was $10.24\,\t{\mu m}$, which is sufficiently large to prevent effects of the reflection form the boundaries of the simulation area. The simulation cell size was $5\times5\times2\,\t{nm}^3$. Since the parametric excitation requires thermal seeds of magnons (thermal magnon population), we enabled thermal fluctuations in the simulations setting the temperature to $T = 1\,$K (please note, that simulation temperature in MuMax3 poorly describes real properties of thermal field, and can be used for qualitative estimations, but not for the quantitative analysis).

In the following, we consider two SW frequencies, $5.2\,$GHz and $7.845\,$GHz, which correspond to the SW wavenumbers $k = 0.0377\,\t{nm}^{-1}$ and $k = 0.0628\,\t{nm}^{-1}$, as was confirmed by the simulation of SW dispersion and analytical calculations \cite{Verba_PRAppl2017}. The SW wavelengths are, thus, $\lambda_\t{SW} = 167\,$nm and  $\lambda_\t{SW} = 100\,$nm, respectively. The chosen wavenumbers ensure that the  parametric pumping is adiabatic , i.e. $\t{sinc}[kL_p] = 0$.

First, we simulated the reference case of a traditional parametric transducer. For this, we assumed rf-modulated effective volume perpendicular anisotropy in the area under the parametric gate,
	\begin{equation}
	   K_u (t) = K_u + K_{rf} \sin[2\pi f_p t + \vp_p] \,,
	\end{equation}
where the modulation, i.e. the pumping strength, is related to the microwave electric field as $E_{rf} = K_{rf} h / \beta$, and the pumping frequency is the double SW frequency, i.e., $f_p = 10.4\,$GHz and $f_p = 15.69\,$GHz for the chosen SW wavenumbers. The simulations were run for 200~ns to reach a steady state. In fact, such a large time is required only very close to the excitation threshold.  For stronger pumping the steady state is achieved much faster. 

In the simulations the  magnetization dynamics was recorded at 250~nm to the right of the gate (in $+x$ direction). By fitting the simulation data with a sine function $m_z(t) = m_z \sin[\pi f_p t + \vp_\t{SW}]$, we were able to extract the SW  amplitude $m_z$ and phase $\vp_\t{SW}$.

\begin{figure}
	\includegraphics[width=\columnwidth]{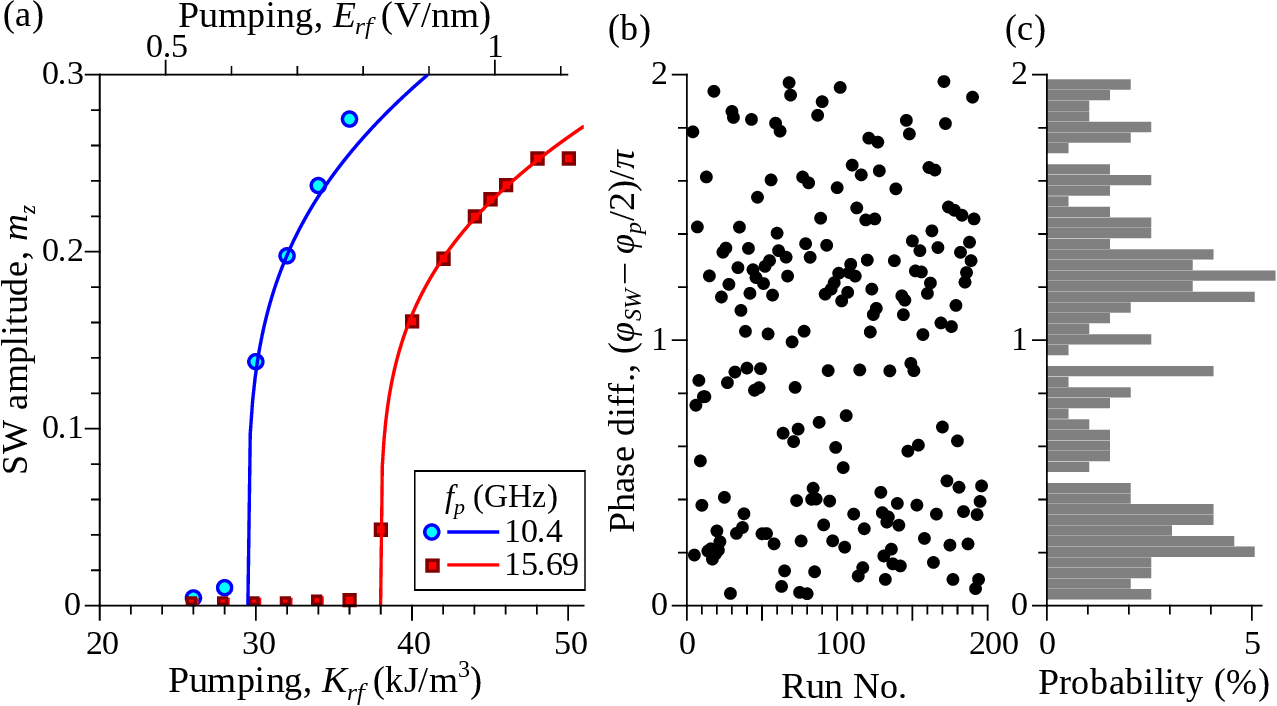}
	\caption{Case of a traditional parametric excitation gate: (a) Amplitude of parametrically excited SW (dimensionless dynamic magnetization component $m_z$ recorded at 250~nm distance from the excitation gate) as a function of the pumping strength $K_{rf}$.  Symbols -- micromagnetic results, lines -- fit by Eq.~(\ref{e:fit}); (b,c) Phase difference between the excited SW and half the pumping phase, $\vp_\t{SW} - \vp_p/2$, at the pumping frequency $f_p = 10.4\,$GHz and pumping strength $K_{rf} = 29\,\t{kJ}/\t{m}^3$: (b) -- raw simulation data of 200 simulation runs, (c) -- histogram of the phase distribution.}\label{f:sim-Vdc0}
\end{figure}

SW amplitudes are shown in Fig.~\ref{f:sim-Vdc0}(a). We see a clear threshold behavior -- amplitude remains low (at thermal level) below the threshold and rapidly increases above it. To quantify the threshold value, the data were fitted by the dependence:
	\begin{equation}\label{e:fit}
		m_z = \mathcal{A} \sqrt[4]{K_{rf}^2 - K_\t{th}^2} \,,
	\end{equation}
which is a standard dependence of SW amplitude above the threshold, if the dominant process limiting the growth of  parametric instability is the four-magnon interaction between the excited magnons \cite{Verba_PRAppl2017, Lvov_Book1994}. The threshold, naturally, increases with the pumping frequency, because higher-frequency SW have a larger group velocity.  The absolute values of the SW amplitudes agree well with the results of theoretical calculations (see comparison below in Fig.~\ref{f:sim-amp}).

Figure \ref{f:sim-Vdc0}(b) shows the difference between the excited spin wave phase and half of the pumping phase $\Delta \vp = \vp_\t{SW} - \vp_p/2$, which has been calculated over 200 independent simulation runs. As was expected for adiabatic parametric pumping, the phase difference is random, and covers all the phase space. The results of the histogram analysis (Fig.~\ref{f:sim-Vdc0}(c)), where the probability to find the phase difference inside one of the $[\pi (n-1)/25, \pi n/25]$, $n=1,\dots,50$ regions is shown, reveals that the phase distribution is not uniform, as certain weakly pronounced peaks at $0.2\pi$ and $1.2\pi$ are visible. We attribute this observation to numerical artifacts (which may be caused by some non-adiabatic contributions due to the discretization), and relatively small size of the used data set, which was limited because of a prohibitively large  computational time. 

In order to quantify the phase distribution by a single number, we recall that the probability to observe the phase difference $\Delta\vp$ and $\Delta\vp + \pi$ should be the same -- this follows from the definition of the $\Delta\vp$ and obvious periodicity of the dynamics with $2\pi$ change of the pumping phase. We thus sum the observations of $\Delta\vp$ and $\Delta\vp + \pi$, mapping thus the distribution from $2\pi$ range into $\pi$ range. The resulting distribution becomes approximately single-peak, allowing to characterize it by the phase variance $\sigma_\vp = \langle \Delta\vp^2 \rangle - \left(\langle\Delta\vp\rangle\right)^2$. For the traditional parametric pumping (Fig.~\ref{f:sim-Vdc0}), we get $\sigma_\vp = 0.07 \pi$, which is close to the value, expected for the uniform probability distribution ($\sigma_\vp/\pi = 1/12 \approx 0.083$).

Next, we simulate the effect of SW  reflection at a transducer lateral boundary by changing the static magnetic anisotropy  $K_u \to K_u + \Delta K_u$ under the control gate on the magnetic waveguide Fig.~\ref{f:sketch}. Modified anisotropy shifts SW dispersion under the control gate, thus creating a wavenumber mismatch for the incoming parametric SW, which, in turn, results in a partial reflection of the parametrically excited  wave. The results are shown in Fig.~\ref{f:sim-amp}. We observe a clear shift of the threshold curves towards lower pumping strengths (panel (a)). Under sufficiently strong dc control field and pumping, simulations show certain decrease in the SW amplitude, which is due to the transition to a more complex parametric dynamics involving more than one magnon pair. 

\begin{figure}
	\includegraphics[width=\columnwidth]{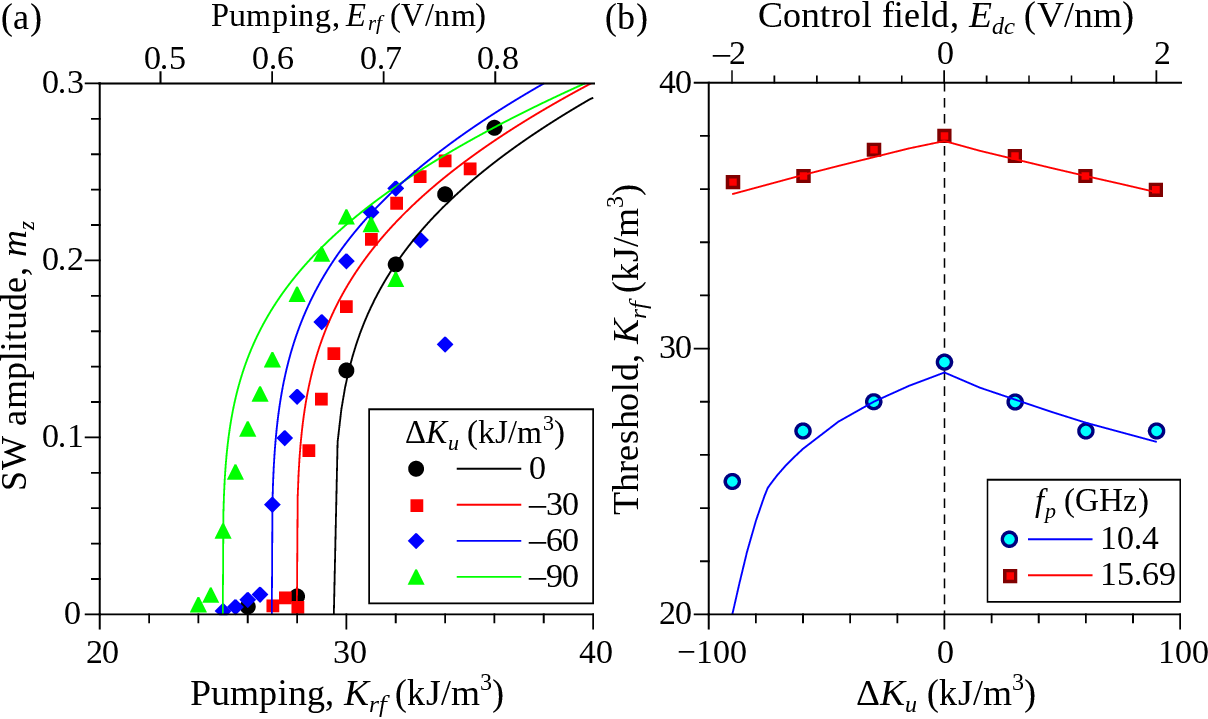}
	\caption{(a) Amplitude ($m_z$ component) of the parametrically excited SW as a function of the pumping strength for different values of the anisotropy variation $\Delta K_u$ under the control gate.  The pumping frequency is $f_p = 10.4\,$GHz; symbols -- simulation, lines -- fits by Eq.~(\ref{e:fit}); (b) Effect of the anisotropy variation on the SW excitation threshold for different pumping frequencies.  Top axis shows corresponding dc electric field at the control gate.  Symbols -- simulations, lines -- theoretical calculations.}\label{f:sim-amp}
\end{figure}

Using the same fitting procedure on the micromagnetic data calculated near the threshold, we extract the threshold of parametric excitation as a function of the anisotropy variation $\Delta K_u$,  and, thus, as a function of the control electric field (Fig.~\ref{f:sim-amp}(b)). This figure shows that the control electric field of both signs leads to the  reduction of the parametric threshold, because any shift (up or down) of the SW dispersion under the control gate produces a nonzero reflection. 

The weaker reduction of the excitation threshold for higher pumping frequency -- $5\%$ for 15.69~GHz as compared to $15\%$ for 10.4~GHz for experimentally feasible control electric field  of $E_\t{dc} = \pm 2\,$V$/$nm -- is explained by a lower SW reflection coefficient under the same dc voltage. Indeed, the higher is the magnon wavenumber, the lower is its sensitivity to anisotropy or static magnetic field (in the exchange limit, $\partial k/\partial K_u \sim 1/k$), as the difference of wavenumbers matters for the reflection rate, see below.

For the comparison of the simulations results with the above developed theory, we used a well-known expression for dispersion relation of the fundamental (uniform) SW mode in the considered geometry of axially magnetized nano-waveguide \cite{Verba_PRAppl2017}:
	\begin{equation}
		\omega_{k} = \sqrt{\Omega_{k}^{(yy)} \Omega_k^{(zz)}} \,,
	\end{equation}
where 
	\begin{equation}
		\begin{split}
		\Omega_k^{(yy)} &= \omega_M \left(\lambda^2 k^2 + F_{k}^{(yy)} \right)\,, \\  \Omega_k^{(zz)} &= \omega_M \left(\lambda^2 k^2 + F_{k}^{(zz)} - n_a \right) \,,
		\end{split}
	\end{equation}
$k = k_x$ is the spin wave wavenumber along the waveguide,  $\omega_M = \gamma \mu_0 M_s$, $\gamma$ is the gyromagnetic ratio, $\lambda = (2 A_\t{ex}/(\mu_0 M_s^2))^{1/2}$ is the exchange length, $\mat F_{k}$ is the dipolar tensor, and $n_a = 2K_u /(\mu_0 M_s^2)$ describes perpendicular magnetic anisotropy. 

The parametric coupling efficiency was derived in \cite{Verba_PRAppl2017}
	\begin{equation}
		V = V_{k_x} = \frac{\gamma}{4} \sqrt{\frac{\Omega_k^{(yy)}}{\Omega_k^{(zz)}}}\ .
	\end{equation}
The relations between the effective pumping field $b_p$, anisotropy modulation amplitude $K_{rf}$ and microwave electric field $E_{rf}$ have been already given above. Finally, for the calculation of the reflection coefficient, we use the expression:
	\begin{equation}\label{e:R}
		R = \frac{k_0-k_1}{k_0 + k_1} \,,
	\end{equation}
which is rigorous for exchange SW,  and in nano-waveguides works reasonably well for dipole-exchange spin waves too \cite{Verba_PRB2020bound}. Here, $k_0$ and $k_1$ are the wavenumbers of the incident (parametric) wave and the wave under the control gate. 

From Fig.~\ref{f:sim-amp}(b) one finds a good quantitative agreement between the theory and simulations. Certain discrepancy appears at large negative control electric field and lower pumping frequency.  This happens because the SW of the given frequency under the control gate enters into a deeply dipolar range (has low $k$), where the accuracy of Eq.~(\ref{e:R}) becomes insufficient.

\begin{figure}
	\includegraphics[width=\columnwidth]{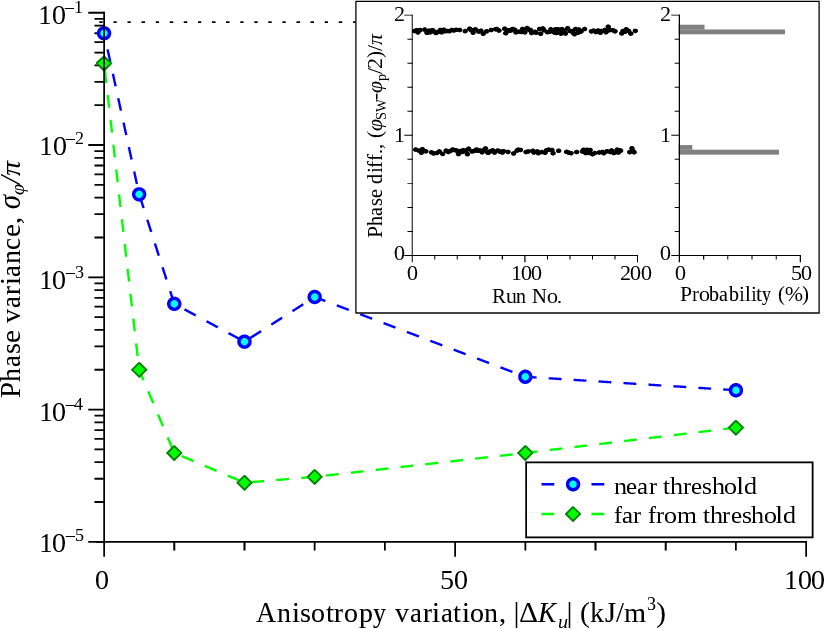}
	\caption{Phase variance of the parametrically excited SW as a function of the anisotropy change $|\Delta K_u|$ (range of negative $\Delta K_u$ is simulated) under the control gate:   micromagnetic results, lines are guide to an eye. Dotted line shows the variance for uniform distribution. The pumping frequency is $10.4\,$GHz, pumping strength was chosen close to the threshold (small-amplitude regime, $m_z \sim 0.05$), and far above the threshold (strongly nonlinear regime with $m_z \approx 0.2$). Inset: example phase distribution across 200 runs and corresponding histogram for $\Delta K_u = -90\,\t{kJ}/\t{m}^3$ and low-amplitude regime ($K_{rf} = 25\,\t{kJ}/\t{m}^3$).}\label{f:sim-phase}
\end{figure}

From the micromagnetic data we also extracted the phase of the excited spin SW (referred to half of the pumping phase). In a great contrast to the case of no control gate (Fig.~\ref{f:sim-Vdc0}(b,c)), in the presence of reflection from the control gate, the phase distribution becomes very narrow, forming two clear peaks separated by $\pi$ radians (see inset in Fig.~\ref{f:sim-phase}). That is exactly what we expected from our theoretical predictions, and what is realized in the case of a parametric oscillator. Under the strongest control electric field (corresponding to $\Delta K_u = -90\,\t{kJ}/\t{m}^3$), the phase variance $\sigma_\vp$ drops by almost three orders of magnitude. 

It is important to note, that even relatively small control electric fields ($\Delta K_u = -5\,\t{kJ}/\t{m}^3$), which leads to only a small reflection coefficient $\tilde R = 0.012$ , and, thus, has a marginal effect on the excitation threshold, causes ten times reduction in the SW phase variance (see Fig.~\ref{f:sim-phase}). Of course, a strong thermal fluctuations may impose a requirement to have a larger reflection coefficient to lead to a substantial narrowing of the phase distribution, but this investigation lies beyond the scope of our current work.

In addition, we verified if the effect of phase distribution narrowing is robust against the SW amplitude, i.e. if it persists in a deeply nonlinear regime. For this, we repeated the simulations, setting the pumping amplitude to the value, corresponding to amplitude of the excited SW of  $m_z \approx 0.2$ at each value of the used reflection coefficient. The main effect was still the same:  at zero reflection the phase variance was large, $\sigma_\vp = 0.041/\pi$ (a somewhat smaller value than in the small-amplitude case). The most probable reason for that can be attributed to a weak pumping non-adiabaticity caused by a nonlinear shift of the SW dispersion.  The increase of the reflection coefficient from the control gate causes a rapid drop of phase variance to below $10^{-4}$ level. The inverted dependence of the phase variance on  $|\Delta K_u|$  observed at large  $|\Delta K_u|$ values in the strongly above-threshold regime  is, most probably, a numerical artifact caused by finite discretization. The same can be said about the origin of the out-of-trend point at $|\Delta K_u| = 30\,\t{kJ}/\t{m}^3$ in the near-threshold regime  (Fig.~\ref{f:sim-phase}). 

In addition to the excitation of short exchange-dominated spin waves with narrow phase distribution, ability to control the SW phase variance by static voltage (or any other means) could be potentially interesting for probabilistic computing. Point-like magnetic parametric oscillator has been already proposed and investigated for the purpose of probabilistic computing in \cite{Makiuchi_APL2021, Hioki_JAP2022}. Such an oscillator possesses only two possible phase states, probability of which can be controlled by a weak lock-in signal. In the case of moderate control fields, our system offers the same properties, but with magnon (instead of electric) output, which can be useful for integrated magnonic circuits \cite{Guo_ArXiv2026}. At low control electric fields, the phase distribution becomes essentially continuous with controlled variance, which could be interesting for other, less common, approaches in probabilistic computing.


\section{Conclusion}\label{s:conclusion} 

In this work, we proposed, proved analytically and verified by micromagnetic simulations a new method of parametric excitation of nano-scale exchange-dominated SW with a controlled narrow phase distribution. The method is based on the utilization of an open-resonator-type transducer, where an additional dc-controlled gate is used for a partial reflection of one of the excited parametric magnons. This partial reflection imposes an additional  phase relation between the parametric magnons, which, together with the phase relation between the parametric magnons and pumping, restrict the possible phase state of the excited magnons to only two allowed phases, similar to the  of a parametric oscillator. Micromagnetic simulations, performed with the account of thermal fluctuations, confirmed a drastic reduction of the phase variance of the excited SW by up to three orders of magnitude, which holds in a deeply nonlinear (high-amplitude) regime as well. 

In contrast to the alternative competing  approach of non-adiabatic pumping, our method does not rely on transducers  of a sub-wavelength width, possessing hardly reachable thresholds or wavelength-selective patterned transducers. The proposed principle is not limited to thin waveguides with VCMA,  but works for any other kind of parametric pumping (e.g., microwave magnetic field) and control of the reflection coefficient (e.g., by dc current), and, thus, is applicable to various magnonics systems based on metallic or dielectric ferromagnets. It also will work for the parametric excitation of waves of a different nature (e.g. acoustic or electromagnetic). Furthermore, the ability to control phase distributions of continuously excited waves by a weak reflection from the excitation transducer boundaries  could be potentially interesting for probabilistic computing, random signal generation and other applications based on stochastic physics.

\section*{Acknowledgements}
R.V. acknowledge support by the National Research Foundation of Ukraine, Grant No. 2025.03/0005.
A.S. and V.T. acknowledge support from the U.S. Army and the Oakland University foundation. 

\bibliography{Refs-VCMA}

\end{document}